\documentclass[aps,preprint,floatfix]{revtex4}%
\usepackage{amsfonts}
\usepackage{amsmath}
\usepackage{amssymb}
\usepackage{graphicx}
\usepackage{graphicx}%
\begin{document}
\title{Increasing efficiency of quantum memory based on atomic frequency combs}
\author{R. N. Shakhmuratov}
\affiliation{Zavoisky Physical-Technical Institute, FRC Kazan Scientific Center of RAS,
Kazan 420029, Russia}

\begin{abstract}
A protocol, which essentially increases the efficiency of the quantum memory
based on the atomic frequency comb (AFC), is proposed. It is well known that a
weak short pulse, transmitted trough a medium with a periodic structure of
absorption peaks separated by transparency windows (AFC), is transformed into
prompt and delayed pulses. Time delay is equal to the inverse value of the
frequency period of the peaks. It is proposed to send the prompt pulse again
through the medium and to make both delayed pulses to interfere. This leads to
the essential increase of the efficiency of the AFC storage protocol.

\end{abstract}
\maketitle

\section{Introduction}

Single photons are ideal information carriers propagating fast a long distance
with low losses. Controlling single photons is an important point in quantum
computing and quantum telecommunication. One of the experimental challenges in
quantum information science is a coherent and reversible light-matter mapping
of quantum information carried by a single-photon wave packet. A light-state
storage in collective atomic excitations for a pre-determined time is one of
the ways to realize quantum memory crucial for quantum repeaters in quantum networks.

There are many schemes of quantum memory employing photon-echo technique
\cite{Chaneliere2009,Tittel2010}, controlled reversible inhomogeneous
broadening (CRIB) protocol \cite{Moiseev2001,Manson2006,Manson2007},
electromagnetically induced transparency \cite{Lukin2000,Lukin2001},
off-resonant Raman interaction \cite{Nunn,Reim2010,Nunn2011}. The list of
methods and related references are far to be exhausted. These methods suffer
from contamination of the signal channel by spontaneous emission caused by the
strong classical fields exciting auxiliary transitions in atoms or as in the
case of CRIB protocol can be realized if optical transitions are sensitive to
the electric fields controlling inhomogeneous broadening.

Passive schemes as, for example, atomic frequency comb (AFC) protocol
\cite{Gisin2008,Gisin2009,Bonarota,Chaneliere,Bonarota2012}, are preferable
since passive protocols are capable to store quantum information in collective
atomic excitations for a pre-determined time without using additional
excitations complicating the storage schemes. Meanwhile, quantum efficiency of
the AFC protocol, which is the first example of the passive scheme, is limited
to $54\%$ \cite{Gisin2009}. A lambda type excitation of AFC on an auxiliary 
transition by strong control fields is capable to increase quantum efficiency of 
the AFC protocol close to $100\%$, see Ref. \cite{Gisin2009}. However, combination 
of the passive AFC scheme with the auxiliary excitation also contaminates the
quantum channel.

In this paper, a modification of the AFC protocol, which helps to improve the
quantum efficiency without using additional fields, is proposed. A short pulse
propagating through a medium with the AFC absorption spectrum is transformed
into a prompt and delayed pulses at the exit of the medium. For the optimal
values of the optical thickness of the medium and finesse of the comb the
intensities (amplitudes) of the prompt and delayed pulses are $0.13I_{0}$
($0.37E_{0}$) and $0.54I_{0}$ ($0.73E_{0}$), respectively, where $I_{0}$ and
$E_{0}$ are maximum intensity and amplitude of the incident pulse. If the
prompt pulse is transmitted again through the same AFC medium a new pulse with
the same delay time is produced. The amplitude of this pulse is $0.27E_{0}$.
If we make two delayed pulses interfere constructively, the intensity of the
produced pulse will be close to $I_{0}$. Physical constrains and limitations
are considered in this paper.

\section{Efficiency of the direct AFC protocol}

In this section, the efficiency of the AFC quantum memory and results,
obtained in Refs. \cite{Gisin2008,Gisin2009,Bonarota,Chaneliere,Bonarota2012}
are analyzed.

Frequency combs consisting of the absorption peaks separated by the
transparency windows can be prepared in an inhomogeneously broadened
absorption line of rare-iron doped crystals by different methods. One of them
employs a long sequence of pulse pairs separated by time $T$, see Refs.
\cite{Gisin2008,Gisin2009,Bonarota,Chaneliere}. Each pair creates a frequency
comb with a period $2\nu_{0}=2\pi/T$ due to pumping ground state atoms to a
long-lived shelving state. Accumulative effect of many pairs of relatively
weak pulses is capable to create deep holes in the inhomogeneously broadened
absorption spectrum. Such a pumping creates a harmonic structure of an atomic
population difference in the spectrum,%
\begin{equation}
n(\Delta)=\frac{1}{2}\left[  1-\cos\left(  \frac{\pi\Delta}{\nu_{0}}\right)
\right]  , \label{Eq1}%
\end{equation}
where $\Delta=\omega_{c}-\omega_{A}$ is a frequency difference of a pulse
carrier, $\omega_{c}$, and individual atom in the comb, $\omega_{A}$, see
Refs. \cite{Saari1994,Shakhmuratov2018}. Here, atoms producing the absorption
peaks occupy at their centers the ground state, while at the bottom of the
transparency windows all atoms are removed to the shelving state.
Inhomogeneous broadening is assumed to be very large. Therefore, the
difference between the absorption peaks is neglected on the frequency scale
comparable with a spectrum of optical pulses, which are filtered by the AFC.

In the other method, one creates a broad transmission hole by spectral hole
burning and then periodic narrow-spectrum ensembles of atoms are created in
the hole by repumping atoms from the storage state to the ground state
\cite{Kroll2005,Kroll2010_2,Kroll2010}. Periodic structure of Loretzians,
created by this method, was considered in Refs.
\cite{Bonarota,Chaneliere,Bonarota2012}. This structure is described by%
\begin{equation}
n(\Delta)=\sum_{k=-N-1}^{N}\frac{\Gamma^{2}}{[\Delta+\nu_{0}(2k+1)]^{2}%
+\Gamma^{2}}, \label{Eq2}%
\end{equation}
where $2\nu_{0}$ is the period of the comb, $\Gamma$ is a halfwidth of the
absorption peaks, and $2N+2$ is a number of peaks. It is also possible to
create AFC with square-shaped absorption peaks of width $2\delta$ separated by
transparency windows. The distance between the centers of the absorption peaks
is equal to $2\nu_{0}$. Then, the distribution of the atomic population
difference in the inhomogeneously broadened absorption spectrum is described
by the function%
\begin{equation}
n(\Delta)=\sum_{n=-N-1}^{N}\left\{  \theta\left[  \Delta-(2k+1)\nu_{0}%
+\delta\right]  -\theta\left[  \Delta-(2k+1)\nu_{0}-\delta\right]  \right\}  ,
\label{Eq3}%
\end{equation}
where $\theta(x)$ is the Heaviside step function. AFC with square-shaped
absorption peaks are prepared in Ref. \cite{Bonarota} by a differen method,
which employs a pulse train with special relations of phases and amplitudes
distributed according to the sinc function. Also, chirped light pulses with
hyperbolic-secant complex amplitudes were used to built square-shaped
absorption peaks in Ref. \cite{Bonarota2012}. Examples of two AFCs with
Lorentzian and square-shaped peaks are shown in Fig. 1. It is instructive to
compare these two AFCs since they have tunable finesse, which is $F_{S}%
=\nu_{0}/\delta$ for the square-shaped AFC and $F_{L}=\nu_{0}/\Gamma$ for the
Lorentzians. Finesse of the harmonic AFC, described by Eq. (\ref{Eq1}), is
fixed and equal to $F_{H}=2$. Therefore, this AFC can be compared with the
other two when $F_{S}=F_{L}=2$. \begin{figure}[ptb]
\resizebox{0.5\textwidth}{!}{\includegraphics{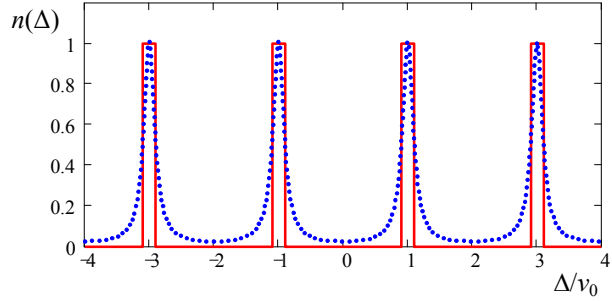}}\caption{Frequency
distribution of the population difference of atoms in the spectrum for AFC
with square-shaped peaks (red solid line) and Lorentzian peaks (blue dotted
line). The distance between the peak centers is $2\nu_{0}$. Half width of the
peaks is $\delta$ for the square-shaped peaks and $\Gamma$ for the Lorentzian
peaks. They are $\delta=\Gamma=\nu_{0}/10$.}%
\label{fig:1}%
\end{figure}

At the exit of the medium with a periodic absorption spectrum, a pulse with a
spectrum covering many (or at least several) absorption peaks of the comb is
transformed into a prompt pulse and several delayed pulses with delay times
equal to $T$, $2T$, $3T\,$, etc., see Refs.
\cite{Gisin2008,Gisin2009,Bonarota,Chaneliere,Bonarota2012,Saari1994,Shakhmuratov2018}%
. All the pulses have the same shape coinciding with the shape of the incident
pulse. For the combs with high finesse, $F\gg2$, and moderate optical
thickness of the absorption peaks, only the prompt and first delayed pulses
have noticeable amplitudes, see Appendix A. Below we focus on the properties
of these two pulses, which are%
\begin{equation}
E_{out}(t)=E_{p}(t)+E_{d}(t-T), \label{Eq4}%
\end{equation}
where $E_{out}(t)$ is the field at the exit of the AFC medium, $E_{p}(t)$ is
the prompt pulse with no delay, and $E_{d}(t-T)$ is the first delayed pulse,
the amplitude of which takes maximum value at time $t=T$, while maximum of the
prompt pulse is localized at $t=0$ by definition.

Maximum amplitude of the prompt pulse $E_{p}(0)=C_{0}E_{0}$, where $E_{0}$ is
a maximum amplitude of the incident pulse, is reduced by the coefficient
$C_{0}$, which is%
\begin{equation}
C_{0}=\left\{
\begin{array}
[c]{cc}%
e^{-\frac{d_{p}}{2F_{H}}} & \mathrm{harmonic}\\
e^{-\frac{\pi d_{p}}{4F_{L}}} & \mathrm{Lorentzians}\\
e^{-\frac{d_{p}}{2F_{S}}} & \mathrm{squares}%
\end{array}
\right.  , \label{Eq5}%
\end{equation}
see Refs. \cite{Bonarota,Chaneliere,Bonarota2012,Shakhmuratov2018} and
Appendix A. Here the label denotes the type of AFC, $d_{p}=\alpha l$ is an
optical thickness of the medium for a monochromatic field tuned in resonance
with one of the absorption peaks, $\alpha$ is the corresponding Beer's law
attenuation coefficient, $l$ is a physical thickness of the medium, and $F$ is
finesse of the comb, which is $F_{H}=2$, $F_{L}=\nu_{0}/\Gamma$ or $F_{S}%
=\nu_{0}/\delta$ depending on the selected AFC.

Maximum amplitude of the first delayed pulse is $E_{d}(0)=C_{1}E_{0}$, where%
\begin{equation}
C_{1}=C_{0}\times\left\{
\begin{array}
[c]{cc}%
\frac{d_{p}}{2F_{H}} & \mathrm{harmonic}\\
\frac{\pi d_{p}}{2F_{L}}e^{-\frac{\pi}{F_{L}}} & \mathrm{Lorentzians}\\
\frac{d_{p}}{\pi}\sin\left(  \frac{\pi}{F_{S}}\right)  & \mathrm{squares}%
\end{array}
\right.  , \label{Eq6}%
\end{equation}
$C_{0}$ is the coefficient in Eq. (\ref{Eq5}), corresponding to the relevant
comb, see Refs. \cite{Bonarota,Chaneliere,Bonarota2012,Shakhmuratov2018} and
Appendix A. The coefficient $C_{1}$ takes global maximum value when
$C_{0}=e^{-1}$. This value is achieved if the optical depth $d_{p}$ is equal
to $4$ for the harmonic comb, $4F_{L}/\pi$ for the comb of Lorentzian peaks,
and $2F_{S}$ for the square comb. For these values of $d_{p}$ and large
finesse satisfying the condition $F_{L,S}\gg\pi$, we have%
\begin{equation}
C_{1}=e^{-1}\times\left\{
\begin{array}
[c]{cc}%
1 & \mathrm{harmonic}\\
2e^{-\frac{\pi}{F_{L}}} & \mathrm{Lorentzians}\\
2 & \mathrm{squares}%
\end{array}
\right.  . \label{Eq7}%
\end{equation}
Extra exponent for the comb of Lorentzians originates from the inhomogeneous
broadening of the absorption peaks with Lorentzian wings, which give
$\exp(-\pi/F_{L})=\exp(-\Gamma T)$. Therefore, the square-shaped comb produces
the first delayed pulse with larger amplitude for moderate values of finesse
($F_{S}>\pi$) compared with the comb consisting of the Lorentzian peaks.

Meanwhile, an abrupt drop of the wings of the square-shaped peaks results in
the function $\sim\sin\left(  \frac{\pi}{F_{S}}\right)  $, see Eq.
(\ref{Eq6}), which also reduces the amplitude of the first delayed pulse
produced by the pulse filtering through this comb with a moderate value of finesse.

The intensity of the first delayed pulse is defined by the equation
$I_{1}=C_{1}^{2}I_{0}$, where $I_{0}$ is a maximum intensity of the incident
pulse. Dependence of the coefficient%
\begin{equation}
C_{1}^{2}=\left(  \frac{d_{p}}{\pi}\right)  ^{2}\sin^{2}\left(  \frac{\pi
}{F_{S}}\right)  e^{-\frac{d_{p}}{F_{S}}} \label{Eq8}%
\end{equation}
for the square comb on $d_{p}$ for different values of the finesse $F_{S}$ is
shown in Fig 2(a). For example, for $F_{S}=2$ the optimal value of the optical
thickness is $d_{p}=4$ and corresponding global maximum of the first delayed
pulse is $I_{1}=0.219I_{0}$. For the harmonic frequency comb, this global
maximum is even smaller and equals to $I_{1}=e^{-2}I_{0}=0.135I_{0}$. With the
increase of $F_{S}$ or with narrowing of the absorption peaks at the fixed
width of the transparency windows, the optimal value of $d_{p}$ also increases
according to the relation $d_{p}=2F_{S}$ and the optimal intensity of the
first delayed pulse, $I_{1}$, corresponding to the global maximum rises, see
Fig. 2(b). For example, for $F_{S}=10$ the optimal value of $d_{p}$ is $20$
and maximum value of $I_{1}$ is $0.524I_{0}$. If large finesse $F_{S}$ does
not correspond to the optimal value of the optical thickness $d_{p}/2$, then
maximum value of $I_{1}$ decreases. For example, for $d_{p}=2$ and $F_{S}=10$,
the maximum amplitude of the first delayed pulse is $I_{1}=0.104I_{0}$, which
is almost five times smaller than that for the optimal values of the
parameters $d_{p}=20$ and $F_{S}=10$.

Maximum efficiency of the AFC quantum memory, $54\%$, is achieved for a very
large values of finesse and optical thickness. For the square-shaped comb this
efficiency is realized when $F_{S}=32$ and $d=64$. \begin{figure}[ptb]
\resizebox{0.4\textwidth}{!}{\includegraphics{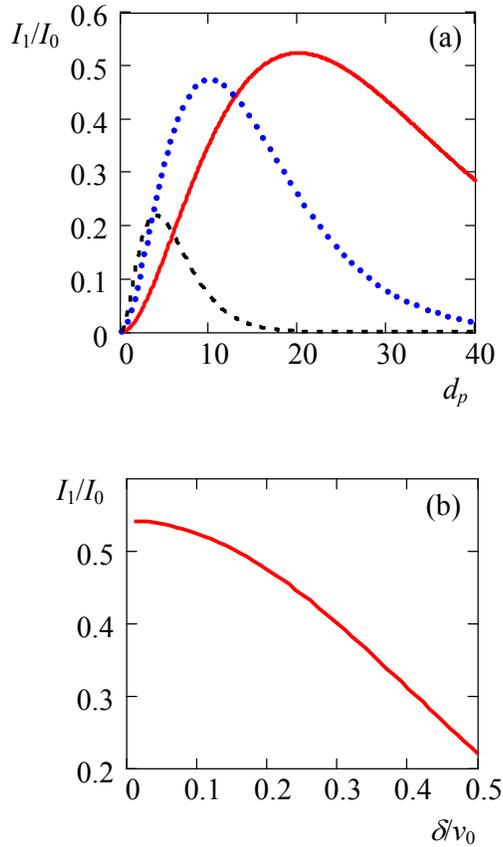}}\caption{(a)
Dependence of the maximum intensity of the first delayed pulse (normalized to
$I_{0}$) on the peak absorption parameter $d_{p}$ for the finesse $F=10$
(solid red line), $5$ (dotted blue line), and $2$ (dashed black line). (b)
Dependence of the intensity $I_{1}/I_{0}$ on the inverse value of the finesse
$1/F=\delta/\nu_{0}$ if $d_{p}=2F_{S}$, which is the condition when this
intensity has a global maximum.}%
\label{fig:2}%
\end{figure}

Actually, it is difficult to create AFC in an optically dense medium since the
hole burning field is strongly absorbed along the sample of large optical
thickness. Therefore, the hole width becomes inhomogeneous along the sample,
i.e. broader at the one side and narrower at the other side. However, this
problem could be solved in a planar geometry where a medium is thick in a
longitudinal direction and relatively thin in a transverse direction. Then,
illuminating along a thin direction with low absorption creates a sequence of
holes, while a weak signal pulse propagating in the longitudinal direction
with high absorption at particular frequencies periodically distributed in a
wide transparency window experiences the necessary splitting into prompt and
delayed pulses. The sequence of holes can be produced by creating a large
spectral hole and then transferring back atoms from an auxiliary state to
create a comb as described in Refs. \cite{Kroll2005,Kroll2010_2,Kroll2010}. If
collinear pulses producing AFC illuminate the sample perpendicular to its thin
side, no spatial grating is created and AFC will be spatially homogeneous
along its thick direction.

Such a geometry was used in Ref. \cite{Rebane1996} to create narrowband
spectral filter, which consists of a planar waveguide, covered with a thin
polymer film containing molecules. They undergo spectral hole burning at
liquid helium temperature creating transparency window at a selected
frequency. Such a scheme of the hole burning was proposed in Ref.
\cite{Shakhmuratov2005} to delay short pulses transmitting them trough an
optically thick sample with a single transparent hole.

\section{Limitations imposed by the homogeneous broadening of the absorption
lines of individual atoms}

In this section the influence of the homogeneous broadening of the absorption
peaks of the comb on the pulse propagation is considered.

To take into account the contribution of the homogeneous broadening we
consider the evolution of the nondiagonal element of the atomic density matrix
$\rho_{eg}(z,t)$ describing coherence between ground $g$ and excited $e$
states of an atom. In the linear response approximation neglecting the change
of the populations $g$ and $e$, the slowly varying complex amplitude of the
atomic coherence, $\sigma_{eg}(z,t)=\rho_{eg}(z,t)\exp(i\omega_{c}t-ik_{c}z)$,
satisfies the equation%
\begin{equation}
\frac{\partial}{\partial t}\sigma_{eg}(z,t)=(i\Delta-\gamma)\sigma
_{eg}(z,t)+i\Omega(z,t)n(\Delta), \label{Eq9}%
\end{equation}
where $\gamma$ is the decay rate of the atomic coherence responsible for the
homogeneous broadening of the absorption line of a single atom, $\Delta
=\omega_{c}-\omega_{A}$ is the difference of the frequency $\omega_{c}$ of the
weak pulsed field and resonant frequency $\omega_{A}$ of an individual atom,
$\Omega(t)=\mu_{eg}E_{0}(z,t)/2\hbar$ is the Rabi frequency, $\mu_{eg}$ is the
dipole-transition matrix element between $g$ and $e$ states, and $n(\Delta)$
is the long-lived population difference, created by the hole burning. Below,
we consider the square-shaped distribution of atoms in the frequency domain,
shown in Fig. 1 by the solid red line. Then, the function $n(\Delta)$ is equal
unity if atom is in the ground state absorbing the resonant field, and
$n(\Delta)$ is zero if atom with the frequency $\omega_{A}=\omega_{c}-\Delta$
is removed by the hole burning to the shelving state resulting in the
appearance of the transparency window.

The Fourier transformation of Eq. (\ref{Eq9}) gives the solution%
\begin{equation}
\sigma_{eg}(z,\nu)=-\frac{\Omega(z,\nu)n(\Delta)}{\nu+\Delta+i\gamma}.
\label{Eq10}%
\end{equation}
With the help of the Fourier transformation of the wave equation%
\begin{equation}
\widehat{L}E_{0}(z,t)=i\hbar\alpha\gamma\left\langle \sigma_{eg}%
(z,t)\right\rangle /\mu_{eg}, \label{Eq11}%
\end{equation}
one can obtain the solution%
\begin{equation}
E_{0}(z,\nu)=E_{0}(0,\nu)\exp[(i\nu z/c)-\alpha_{\mathrm{av}}(\nu)z/2],
\label{Eq12}%
\end{equation}
where $L=\partial_{z}+c^{-1}\partial_{t}$, $\alpha=4\pi\omega_{c}N\left\vert
\mu_{eg}\right\vert ^{2}/\gamma\hbar c$ is the absorption coefficient before
the hole burning, $N$ is the density of atoms, $\left\langle \sigma
_{eg}(z,t)\right\rangle $ is the atomic coherence integrated over
inhomogeneous broadening with the width $\Gamma_{\mathrm{inh}}$, and%
\begin{equation}
\alpha_{\mathrm{av}}(\nu)=i\frac{\alpha\gamma}{\pi\Gamma_{\mathrm{inh}}}%
{\displaystyle\int\nolimits_{-\infty}^{+\infty}}
\frac{n(\Delta)}{\nu+\Delta+i\gamma}d\Delta. \label{Eq13}%
\end{equation}

If inhomogeneous broadening is large enough that the absorption peaks of the
AFC can be considered as having the same height over the frequency range
covered by the spectrum of the incident pulse, then the complex coefficient
$\alpha_{\mathrm{av}}(\nu)$ is reduced to%
\begin{equation}
\alpha_{\mathrm{av}}(\nu)=\alpha\lbrack\epsilon^{\prime\prime}(\nu
)-i\epsilon^{\prime}(\nu)], \label{Eq14}%
\end{equation}
where%
\begin{equation}
\epsilon^{\prime\prime}(\nu)=\frac{1}{\pi}\sum_{k=-N-1}^{N}\left\{  \tan
^{-1}\left[  \frac{\nu+\delta+(2k+1)\nu_{0}}{\gamma}\right]  -\tan^{-1}\left[
\frac{\nu-\delta+(2k+1)\nu_{0}}{\gamma}\right]  \right\}  , \label{Eq15}%
\end{equation}%
\begin{equation}
\epsilon^{\prime}(\nu)=-\frac{1}{2\pi}\sum_{k=-N-1}^{N}\ln\left[  \frac
{[\nu+\delta+(2k+1)\nu_{0}]^{2}+\gamma^{2}}{[\nu-\delta+(2k+1)\nu_{0}%
]^{2}+\gamma^{2}}\right]  . \label{Eq16}%
\end{equation}
Here, $2N+2$ is the number of the absorption peaks in the comb. The plots of
the functions $\epsilon^{\prime\prime}(\nu)$ and $\epsilon^{\prime}(\nu)$ are
shown in Fig. 3. It is seen that the edges of the absorption peaks are
smoothened due to the contribution of the Lorentzian in Eq. (\ref{Eq13})
originating from the response of atoms with population difference $n(\Delta)$
neighboring the frequency component $\nu$. The wings of the Lorentzians also
contribute to the absorption at the centers of the transparency windows where,
for example, at $\nu=0$ we have%
\begin{equation}
\epsilon^{\prime\prime}(0)\approx\frac{4}{\pi}\sum_{k=0}^{N}\frac{\gamma
\delta}{(2k+1)^{2}\nu_{0}^{2}-\delta^{2}}. \label{Eq17}%
\end{equation}
If finesse $F$ is $10$, and $\gamma/\delta=0.1$, then due to the contribution
of the Lorentzians, the absorption at the center of the transparency window
rises from zero to $1.54\times10^{-4}$. For the optimal value of the thickness
$d_{p}$, which is $20$ for $F_{S}=10$, the intensity of a monochromatic
radiation field tuned at the center of the transparency window is reduced by a
factor of $\exp[-d_{p}\epsilon^{\prime\prime}(0)]=0.97$, i.e., it drops by
$3\%$. \begin{figure}[ptb]
\resizebox{0.5\textwidth}{!}{\includegraphics{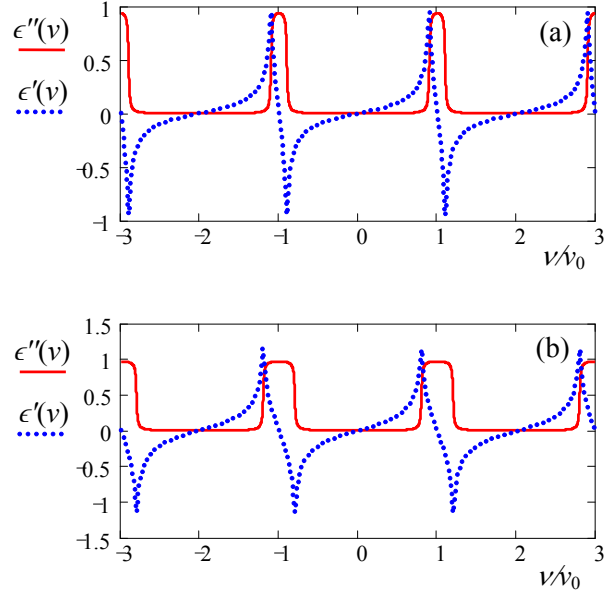}}\caption{(a)
Absorption, $\epsilon^{\prime\prime}(\nu)$, (red solid line) and dispersion,
$\epsilon^{\prime}(\nu)$, (blue dotted line) of the AFC, convoluted with
Lorentzians, see Eqs. (\ref{Eq13})-(\ref{Eq16}). Parameters of the comb are
$\delta/\nu_{0}=0.1$ (a) and $0.2$ (b). Decay rate of the atomic coherence is
the same in both plots, which is $\gamma=0.01\nu_{0}$.}%
\label{fig:3}%
\end{figure}Moreover, maximum of the absorption peaks decreases due to the
homogeneous broadening. For example, the coefficient $\epsilon^{\prime\prime
}(\nu)$ at $\nu=\nu_{0}$ decreases as%
\begin{equation}
\epsilon^{\prime\prime}(\nu_{0})\approx1-\frac{2}{\pi}\left(  \frac{\gamma
}{\delta}-\frac{2\gamma\delta}{4\nu_{0}^{2}-\delta^{2}}\right)  . \label{Eq18}%
\end{equation}
The drop of absorption is noticeable. For the same example, considered for the
transparency windows ($F_{S}=10$ and $\gamma/\delta=0.1$), we have
$\epsilon^{\prime\prime}(\nu_{0})=0.937$, i.e. the absorption coefficient
drops by $6.3\%$ at the centers of the absorption peaks. To reduce this drop
one has to decrease the ratio $\gamma/\delta$ increasing the values of
$\delta$ and $\nu_{0}$.

From the solution, Eq. (\ref{Eq12}), one finds that the prompt pulse and the
first delayed pulse are described by equation%
\begin{equation}
E_{\text{out}}(t)=e^{-A_{0}d_{p}/2}\left[  E_{\text{in}}\left(  t\right)
+A_{1}\frac{d_{p}}{2}E_{\text{in}}\left(  t-T\right)  \right]  , \label{Eq19}%
\end{equation}
where small term $i\nu l/c$ is neglected, and%
\begin{equation}
A_{0}=\frac{1}{2\nu_{0}}\int_{-\nu_{0}}^{\nu_{0}}\epsilon^{\prime\prime}%
(\nu)d\nu, \label{Eq20}%
\end{equation}%
\begin{equation}
A_{1}=-\frac{1}{2\nu_{0}}\int_{-\nu_{0}}^{\nu_{0}}\left[  \epsilon
^{\prime\prime}(\nu)-i\epsilon^{\prime}(\nu)\right]  e^{-i\pi\nu/\nu_{0}}d\nu.
\label{Eq21}%
\end{equation}
From the Kramers-Kronig relations, it follows that the coefficient $A_{1}$ is
reduced to%
\begin{equation}
A_{1}=-\frac{1}{\nu_{0}}\int_{-\nu_{0}}^{\nu_{0}}\epsilon^{\prime\prime}%
(\nu)e^{-i\pi\nu/\nu_{0}}d\nu. \label{Eq22}%
\end{equation}

In spite of the difference between the complex coefficient $\alpha
_{\mathrm{av}}(\nu)=\alpha\lbrack\epsilon^{\prime\prime}(\nu)-i\epsilon
^{\prime}(\nu)]$, averaged with Lorentzian, Eq. (\ref{Eq13}), and the complex
coefficient $\alpha_{c}(\nu)=\alpha\lbrack\chi^{\prime\prime}(\nu
)-i\chi^{\prime}(\nu)]/\chi_{M}$, which is not averaged, see Eq. (\ref{EqA10})
in the Appendix A, the coefficients reducing the absorption of the prompt
pulse, which are proportional to $A_{0}$, Eq. (\ref{Eq20}), for the first
function and $\alpha_{0}/\alpha$, Eq. (\ref{EqA13}), for the second function,
are the same, i.e., $\alpha_{0}/\alpha=A_{0}=\delta/\nu_{0}$. These
coefficients equal to the inverse value of the finesse $F_{S}^{-1}=\delta
/\nu_{0}$, which is the same for both combs. This is almost obvious result.
However, as it was mentioned above, the heights of the absorption peaks and
the depths of the transmission windows of these combs are different and one
could expect that the values of the integrals $A_{0}$ in Eq. (\ref{Eq20}) and
$\alpha_{0}/\alpha$ in Eq. (\ref{EqA13}) responsible for the decrease of the
amplitude of the prompt pulse are also different. Dependencies of $\alpha
_{0}/\alpha=\delta/\nu_{0}$ and numerically calculated $A_{0}$ on $\delta$ for
the fixed values of $\nu_{0}$ and $\gamma$ are compared in Fig 4 demonstrating
that the above conclusion about the same relation of the coefficients with the
finesse, based on the analytical calculation, is correct. Moreover, the
dependence $A_{0}=\delta/\nu_{0}$ is still valid if $\gamma=0.1\nu_{0}$, i.e.,
when the coherence decay rate $\gamma$ is an order of magnitude larger than in
the previous example. \begin{figure}[ptb]
\resizebox{0.4\textwidth}{!}{\includegraphics{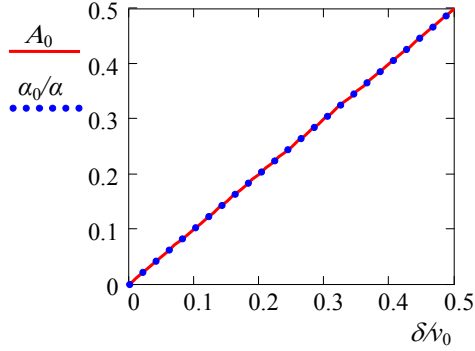}}\caption{Dependencies
of the coefficient $\alpha_{0}/\alpha$ (dotted blue line) and numerically
calculated coefficient $A_{0}$ (red solid line) on $\delta/\nu_{0}$ for a
fixed value of $\nu_{0}$. Homogeneous decay rate of the coherence is
$\gamma=0.01\nu_{0}$.}%
\label{fig:4}%
\end{figure}

The coefficients $A_{1}$ and $-\alpha_{1}/\alpha\sim a_{1}$\ in the solutions
Eq. (\ref{Eq19}) and Eq. (\ref{EqA15}), respectively, which define the
amplitude of the first delayed pulse, are also very close to each other, i.e.,
$-\alpha_{1}/\alpha=2\sin\left(  \frac{\pi\delta}{\nu_{0}}\right)  /\pi$ and
\begin{equation}
A_{1}=2\frac{\sin\left(  \frac{\pi\delta}{\nu_{0}}\right)  }{\pi}e^{-\frac
{\pi\gamma}{\nu_{0}}}. \label{Eq24}%
\end{equation}
The exponential factor $\exp(-\pi\gamma/\nu_{0})=\exp(-T/T_{2})$, where
$T_{2}=1/\gamma$, is the homogeneous dephasing time, has little influence on
the amplitude of the first delayed pulse if $T\ll T_{2}$. It can be shown, see
Ref. \cite{Shakhmuratov2018}, that exactly the same factor, $\exp(-T/T_{2})$,
appears due to the homogeneous dephasing in the expression for the amplitude
of the first delayed pulse $C_{1}E_{0}$, see Eq. (\ref{Eq6}), for the AFSs
consisting of harmonic and Lorentzian peaks.

Experimental verifications of the AFC storage protocol were performed in
Nd$^{3+}$ ions, doped into YVO$_{4}$, Ref. \cite{Gisin2008}, and YAG
(Tm$^{3+}$: YAG)), \cite{Bonarota,Chaneliere,Bonarota2012}. Relatively large
efficiency (9 - 18 $\%$) in Refs. \cite{Bonarota,Chaneliere,Bonarota2012} was
achieved for the moderate value of the initial absorption (before pumping)
described by the parameter $d_{p}\sim4-5$. After the hole burning this
parameter reduced to $d_{p}\sim3$, see Ref.
\cite{Bonarota,Chaneliere,Bonarota2012}. The best performance of this memory
is achieved for the square-shaped AFC \cite{Bonarota,Bonarota2012} with
different values of finesse $F_{S}=2$, $3$, and $5$.

If we take the following values of the AFC parameters realized in Ref.
\cite{Bonarota2012}, i.e., $2\nu_{0}=2$ MHz, $2\delta=400$ kHz, and $\gamma=5$
kHz, then the maximum intensity of the first delayed pulse,%
\begin{equation}
I_{1}=\left(  \frac{d_{p}}{\pi}\right)  ^{2}\sin^{2}\left(  \frac{\pi}{F_{S}%
}\right)  e^{-\frac{d_{p}}{F_{S}}-2\gamma T}I_{0}, \label{Eq25}%
\end{equation}
is nearly $17\%$ of the incident pulse. Finesse of such a comb is $F_{S}=5$.
Actually, the optimal thickness of the sample should be $d_{p}=2F_{S}=10$,
which is only three times larger than that ($d_{p}\sim3$), realized in the
experiments \cite{Bonarota,Chaneliere,Bonarota2012}. For $d_{p}=10$, which is
optimal in this case, the efficiency increases to $46\%$.

\section{Second treatment of the prompt and delayed pulses}

If we split the optical paths of the prompt and first delayed pulses by
time-division multiplexing and send the prompt pulse again through the AFC
medium with the same parameters or backward through the same AFC, we transform
the prompt pulse into new pair of pulses, i.e., the prompt and delayed. This
pair is described by the equation%
\begin{equation}
E_{\text{out}}(t)=e^{-A_{0}d_{p}}\left[  E_{\text{in}}\left(  t\right)
+A_{1}\frac{d_{p}}{2}E_{\text{in}}\left(  t-T\right)  \right]  ,\label{Eq26}%
\end{equation}
where exponential factor $e^{-A_{0}d_{p}}$ differs from that, $e^{-A_{0}%
d_{p}/2}$, in Eq. (\ref{Eq19}) due to absorption in the second path through
the AFC medium. Then, one can make the paths of two delayed pulses such that
both pulses arrive to the selected point at the same time and with the same
phase and then travel together. Their sum is described by the equation%
\begin{equation}
E_{\text{d}}(t-T)=A_{1}\frac{d_{p}}{2}\left(  e^{-A_{0}d_{p}/2}+e^{-A_{0}%
d_{p}}\right)  E_{\text{in}}\left(  t-T\right)  ,\label{Eq27}%
\end{equation}
where small delay time of the pulses due to traveling trough the optical paths
of some length $L$ with a speed of light $c$ is disregarded. Due to
constructive interference of the fields the intensity of this sum field is%
\begin{equation}
I_{\text{d}}(t-T)=\left(  \frac{A_{1}d_{p}}{2}\right)  ^{2}e^{-A_{0}d_{p}%
}\left(  1+e^{-A_{0}d_{p}/2}\right)  ^{2}I_{0}.\label{Eq28}%
\end{equation}
If we take the following optimal values of the parameters for the
square-shaped AFC, considered at the end of Sec. III, i.e., $F_{S}=5$,
$2\nu_{0}=2$ MHz, $\gamma=5$ kHz , and $d_{p}=2F_{S}=10$, then the intensity
of the field $I_{d}(0)$ increases to $86\%$ of the intensity of the incident
pulse. The medium with $d_{p}=2F_{S}=20$ gives even better efficiency, which
is $95\%$. Further increase of the efficiency of the AFC quantum memory is
possible by increasing the frequency spacing of the comb, $2\nu_{0}$, with
respect to the homogeneous width $\gamma$, or by choosing a medium with
smaller value of $\gamma$.

This protocol of quantum storage can be applied to store time-bin qubits,
which in the simplest case can be described as%
\begin{equation}
\left\vert \Psi\right\rangle =C_{1}\left\vert \Psi_{1}\right\rangle
+e^{i\varphi}C_{2}\left\vert \Psi_{2}\right\rangle . \label{Eq29}%
\end{equation}
Here $\left\vert \Psi\right\rangle $ is a single photon state, which is a
superposition of states $\left\vert \Psi_{1}\right\rangle $ and $\left\vert
\Psi_{2}\right\rangle $ corresponding to two short pulses separated in time
and forming time bins, see Refs. \cite{Brendel,Gisin2002}. Quantum information
is encoded in the probability amplitudes $C_{1}$, $C_{2}$ and their relative
phase $\varphi$. These states can be prepared with an unbalance Mach-Zehnder
interferometer, see Refs. \cite{Brendel,Gisin2002} for details. Partial
readouts of the states can be implemented by the same unbalanced Mach-Zehnder
interferometer. Meanwhile, these readouts can be performed using a double-AFC
structure with the frequency periods $2\nu_{1}$ and $2\nu_{2}$, see Refs.
\cite{Gisin2008,Usmani2010}. If the time, $\tau$, between pulses in the
time-bin qubit matches the time difference in delay $\pi(1/\nu_{1}-1/\nu_{2}%
)$, the re-emission from the AFC filters can be suppressed or enhanced
depending on the phase $\varphi$. We do not consider these combined AFC
filters in the storage stage.

We consider the case when time interval between pulses is $\tau<T$ and they
have Gaussian envelopes $\propto e^{-\sigma^{2}(t\pm\tau/2)^{2}}$, with
$+\tau$ for $\left\vert \Psi_{1}\right\rangle $ and $-\tau$ for $\left\vert
\Psi_{2}\right\rangle $. The transformation of these states after passing
through the square AFC is discussed in the Appendix B. Experimental storage
and retrieval of multiple photonic qubits (qudits) consisting of the train of
many pulses is demonstrated in Refs. \cite{Gisin2008,Usmani2010}.

If the train consists of two pulses, after passing through the square AFC the
state $\left\vert \Psi\right\rangle $ is transformed as%
\begin{equation}
\left\vert \Psi\right\rangle _{\mathrm{tr}}=C_{1p}\left\vert \Psi
_{1}\right\rangle _{p}+e^{i\varphi}C_{2p}\left\vert \Psi_{2}\right\rangle
_{p}+C_{1d}\left\vert \Psi_{1}\right\rangle _{d}+e^{i\varphi}C_{2d}\left\vert
\Psi_{2}\right\rangle _{d}, \label{Eq30}%
\end{equation}
where $\left\vert \Psi_{1,2}\right\rangle _{p}$ is a couple of photon states
with no delay (prompt pulses) and $\left\vert \Psi_{1,2}\right\rangle _{d}$ is
a couple of states (actually wave packets) delayed by time $T$. The
coefficients in Eq. (\ref{Eq30}) are $C_{1p}=C_{1}e^{-A_{0}d_{p}/2}$,
$C_{2p}=C_{2}e^{-A_{0}d_{p}/2}$ and $C_{1d}=C_{1p}A_{1}d_{p}/2$,
$C_{2d}=C_{2p}A_{1}d_{p}/2$. The phase factor $e^{i\varphi}$ and relation
between the probability amplitudes of the states $\left\vert \Psi
_{1,2}\right\rangle _{d}$ are the same as for the initial state $\left\vert
\Psi\right\rangle $, Eq. (\ref{Eq29}). If delay time $T$ is much longer than
time separation $\tau$ between pulses in the qubit, the prompt pulses are well
separated from the delayed pulses. Sending the couple of prompt pulses again
through the AFC and making constructive interference of the delayed pulses
from both paths we obtain%
\begin{equation}
\left\vert \Psi\right\rangle _{\mathrm{int}}=\mathcal{C}_{1p}\left\vert
\Psi_{1}\right\rangle _{p}+e^{i\varphi}\mathcal{C}_{2p}\left\vert \Psi
_{2}\right\rangle _{p}+\mathcal{C}_{1d}\left\vert \Psi_{1}\right\rangle
_{d}+e^{i\varphi}\mathcal{C}_{2d}\left\vert \Psi_{2}\right\rangle _{d},
\label{Eq31}%
\end{equation}
where $\mathcal{C}_{1p}=C_{1}e^{-A_{0}d_{p}}$, $\mathcal{C}_{2p}%
=C_{2}e^{-A_{0}d_{p}}$ and $\mathcal{C}_{1d}=C_{1p}(1+e^{-A_{0}d_{p}/2}%
)A_{1}d_{p}/2$, $C_{2d}=C_{2p}(1+e^{-A_{0}d_{p}/2})A_{1}d_{p}/2$. The
probabilities of the delayed pulses are described by equations%
\begin{equation}
\left\vert \mathcal{C}_{1d}\right\vert ^{2}=\left(  \frac{A_{1}d_{p}}%
{2}\right)  ^{2}e^{-A_{0}d_{p}}\left(  1+e^{-A_{0}d_{p}/2}\right)
^{2}\left\vert C_{1}\right\vert ^{2}, \label{Eq32}%
\end{equation}%
\begin{equation}
\left\vert \mathcal{C}_{1d}\right\vert ^{2}=\left(  \frac{A_{1}d_{p}}%
{2}\right)  ^{2}e^{-A_{0}d_{p}}\left(  1+e^{-A_{0}d_{p}/2}\right)
^{2}\left\vert C_{2}\right\vert ^{2}. \label{Eq33}%
\end{equation}
Their forms are exactly the same as for the intensity of the classical field
in Eq.(\ref{Eq28}). Therefore, the conclusion made about efficiency increasing
of the modified AFC memory is also valid for the time-bin quantum states.

\section{Conclusion}

The propagation of the light pulse in a medium with the periodic structure in
the absorption spectrum is analyzed. It is shown that AFCs with the
square-shaped absorption peaks in the spectrum demonstrate larger efficiency
of the field storage. The influence of the homogeneous decay of the atomic
coherence on the quantum efficiency is considered. It is proposed to send the
prompt pulse, transmitted through the AFC medium, to the same medium again and
to make interfere two delayed pulses, i.e., the delayed pulse transmitted
trough the first AFC with the delayed pulse transmitted trough the second AFC.
It is shown that for the optimal parameters of the AFC filters, one can
increase the intensity of the delayed pulse close to the intensity of the
pulse to be stored.

\section{Acknowledgements}

This work was funded by the government assignment from the Federal Research
Center ``Kazan Scientific Center of the Russian Academy of Sciences.''

\section{Appendix A}

In this Appendix the propagation of a short pulse through a medium with the
square AFC in its spectrum is considered. The wave equation, describing the
propagation of the pulsed field $E(z,t)=E_{0}(z,t)\exp(-i\omega_{c}t+ikz)$
along axis $\mathbf{z}$, is (see, for example, Ref. \cite{Crisp})%
\begin{equation}
\left(  \frac{\partial}{\partial z}+\frac{n}{c}\frac{\partial}{\partial
t}\right)  E_{0}(z,t)=i\frac{2\pi\omega_{c}}{nc}P_{0}(z,t), \label{EqA1}%
\end{equation}
where $E_{0}(z,t)$ is the pulse envelope, $k$ is the wave number, $n$ is the
index of refraction, and $P(z,t)=P_{0}(z,t)\exp(-i\omega_{c}t+ikz)$ is the
polarization induced in the medium.

The Fourier transform,
\begin{equation}
F(\nu)=\int_{-\infty}^{+\infty}f(t)e^{i\nu t}dt, \label{EqA2}%
\end{equation}
of the field and polarization satisfy the relation $P(z,\nu)=\varepsilon
_{0}\chi(\nu)E_{0}(z,\nu)$, where $\nu=\omega_{c}-\omega$ is the frequency
difference between the central frequency of the light pulse $\omega_{c}$ and
its spectral component $\omega$, $\varepsilon_{0}$ is the electric
permittivity of free space (below we set $\varepsilon_{0}=1$ and $n=1$ for
simplicity) and $\chi(\nu)$ is the electric susceptibility, which is
\begin{equation}
\chi(\nu)=\chi^{\prime}(\nu)+i\chi^{\prime\prime}(\nu). \label{EqA3}%
\end{equation}
By the Fourier transform the wave equation, Eq. (\ref{EqA1}), is reduced to a
one-dimensional differential equation, the solution of which is (see Ref.
\cite{Crisp})%
\begin{equation}
E_{0}(z,\nu)=E_{0}(0,\nu)\exp\left\{  i\nu\frac{z}{c}-\frac{\alpha z}%
{2\chi_{M}}\left[  \chi^{\prime\prime}(\nu)-i\chi^{\prime}(\nu)\right]
\right\}  , \label{EqA4}%
\end{equation}
where $E_{0}(0,\nu)=E_{in}(\nu)$ is a spectral component of the field incident
to the medium, $\alpha$ is the Beer's law attenuation coefficient describing
absorption of a monochromatic field tuned in maximum of the absorption line
where $\chi^{\prime\prime}(0)=\chi_{M}$ and $\chi^{\prime}(0)=0$. One can
introduce a frequency dependent complex coefficient
\begin{equation}
\alpha_{c}(\nu)=\alpha\frac{\chi^{\prime\prime}(\nu)-i\chi^{\prime}(\nu)}%
{\chi_{M}}, \label{EqA5}%
\end{equation}
which takes into account the contributions of absorption and dispersion.

The imaginary part of susceptibility, $\chi^{\prime\prime}(\nu)$, describes
the field absorption. We consider AFC with the square-shaped absorption peaks
of width $2\delta$ separated by transparency windows. The distance between the
centers of the absorption peaks is equal to $2\nu_{0}$. To make simple
analytical treatment we take Fourier transform of this periodic structure and
limit our consideration to the $2N+1$ spectral components. Then, $\chi
^{\prime\prime}(\nu)$ is expressed as follows%
\begin{equation}
\chi^{\prime\prime}(\nu)/\chi_{M}=\frac{\delta}{\nu_{0}}+\frac{2}{\pi}%
\sum_{k=1}^{N}(-1)^{k}\frac{\sin\left(  \frac{k\pi\delta}{\nu_{0}}\right)
\cos\left(  \frac{k\pi\nu}{\nu_{0}}\right)  }{k}. \label{EqA6}%
\end{equation}
Here $k$ is an integer and inhomogeneous width of the absorption line, where
the frequency comb is prepared, is approximated as infinite. The central
frequency of the comb, $\nu=0$, coincides with the center of one of the
transparency windows corresponding to no absorption in the ideal case. The
height of the absorption peaks corresponds to the absorption of the medium
before the comb preparation.

The real part of the susceptibility, responsible for a group velocity
dispersion, satisfies one of the Kramers-Kronig relations
\begin{equation}
\chi^{\prime}(\nu)=\frac{1}{\pi}\mathcal{P}%
{\displaystyle\int\nolimits_{-\infty}^{\infty}}
\frac{\chi^{\prime\prime}(\nu^{\prime})}{\nu^{\prime}-\nu}d\nu^{\prime},
\label{EqA7}%
\end{equation}
where $\mathcal{P}$ denotes the Cauchy principal value. Calculating the
integral, we obtain
\begin{equation}
\chi^{\prime}(\nu)/\chi_{M}=-\frac{2}{\pi}\sum_{k=1}^{N}(-1)^{k}\frac
{\sin\left(  \frac{k\pi\delta}{\nu_{0}}\right)  \sin\left(  \frac{k\pi\nu}%
{\nu_{0}}\right)  }{k}. \label{EqA8}%
\end{equation}
Frequency dependencies of the absorption $\sim\chi^{\prime\prime}(\nu)$ and
dispersion $\sim\chi^{\prime}(\nu)$\ of the medium with the selected periodic
spectrum are shown in Fig. 5.

Thus, the Fourier transform of the solution of the wave equation (\ref{EqA1}),%
\begin{equation}
E_{0}(z,\nu)=E_{0}(0,\nu)\exp\left[  -\alpha_{c}(\nu)z/2\right]  ,
\label{EqA9}%
\end{equation}
where small term $i\nu z/c$ is neglected, contains the complex coefficient%
\begin{equation}
\alpha_{c}(\nu)/\alpha=\frac{\delta}{\nu_{0}}+\frac{2}{\pi}\sum_{k=1}%
^{N}(-1)^{k}\frac{\sin\left(  \frac{k\pi\delta}{\nu_{0}}\right)  }{k}%
e^{ik\pi\nu/\nu_{0}}, \label{EqA10}%
\end{equation}
which takes into account absorption and dispersion. Their contributions to the
harmonics $e^{ik\pi\nu/\nu_{0}}$ are equal, while the central part,
$\delta/\nu_{0}$, originates only from the absorption. \begin{figure}[ptb]
\resizebox{0.6\textwidth}{!}{\includegraphics{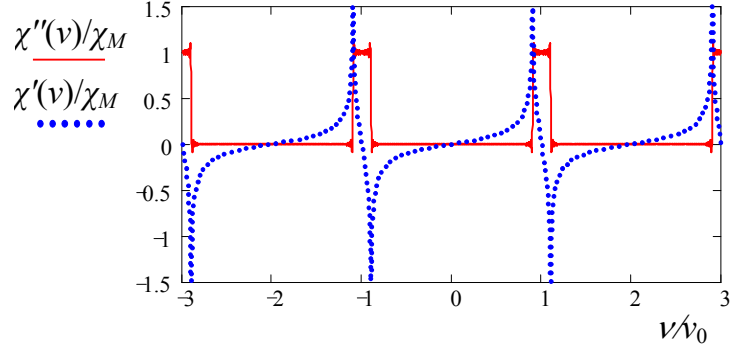}}\caption{ Absorption,
$\chi^{\prime\prime}(\nu)$, (red solid line) and dispersion, $\chi^{\prime
}(\nu)$, (blue dotted line) components of the AFC with square-shaped
absorption peaks of width $2\delta=\nu_{0}/5$ separated by the transparency
windows. Both functions are normalized to $\chi_{0}(\nu)$. Frequency scale is
in units of $\nu_{0}$.}%
\label{fig:5}%
\end{figure}

Similar dependence of $\alpha_{c}(\nu)$ can be derived for any shape of the
absorption peaks since in general for a periodic function $\alpha_{c}(\nu)$ we
have
\begin{equation}
\alpha_{c}(\nu)=\sum_{k=0}^{\infty}\alpha_{k}e^{ik\pi\nu/\nu_{0}},
\label{EqA11}%
\end{equation}
where%
\begin{equation}
\alpha_{k}=\frac{1}{2\nu_{0}}\int_{-\nu_{0}}^{\nu_{0}}\alpha_{c}(\nu
)e^{-ik\nu}d\nu. \label{EqA12}%
\end{equation}
Equations (\ref{EqA10}) and (\ref{EqA11}) contain only positive $k$ due to the
Kramers-Kronig relations. These relations also allow essential simplification
of the expression for $\alpha_{k}$, which can be reduced to%
\begin{equation}
\alpha_{0}=\frac{\alpha}{2\nu_{0}}\int_{-\nu_{0}}^{\nu_{0}}\frac{\chi
^{\prime\prime}(\nu)}{\chi_{M}}d\nu, \label{EqA13}%
\end{equation}
and%
\begin{equation}
\alpha_{k}=\frac{\alpha}{\nu_{0}}\int_{-\nu_{0}}^{\nu_{0}}\frac{\chi
^{\prime\prime}(\nu)}{\chi_{M}}e^{-ik\nu}d\nu\label{EqA14}%
\end{equation}
for $k>0$. In the coefficient $\alpha_{0}$ the dispersion contribution is zero
since it is odd function, while in $\alpha_{k}$ ($k>0$) dispersion,
$\chi^{\prime}(\nu)$, contributes exactly the same value as the absorption,
$\chi^{\prime\prime}(\nu)$. For the same reason the exponents $e^{ik\pi\nu
/\nu_{0}}$ with negative $k$ are absent in Eq. (\ref{EqA11}) since for them
the contributions of $\chi^{\prime}(\nu)$ and $\chi^{\prime\prime}(\nu)$ are canceled.

With the help of the expansion of the function $\exp\left[  -\alpha_{c}%
(\nu)z\right]  $ in a power series of $\exp(i\pi\nu/\nu_{0})$ one finds
\begin{equation}
E_{0}(l,\nu)=E_{in}(\nu)e^{-d_{p}/2F_{S}}\sum_{k=0}^{+\infty}a_{k}e^{i\pi
k\nu/\nu_{0}}, \label{EqA15}%
\end{equation}
where $d_{p}=\alpha_{M}l$, $l$ is the physical length of the medium, $a_{0}%
=1$, $a_{1}=\frac{d_{p}}{\pi}\sin\left(  \pi/F_{S}\right)  $,%
\begin{equation}
a_{2}=-\frac{d_{p}}{2\pi}\sin\left(  \frac{2\pi}{F_{S}}\right)  +\frac
{d_{p}^{2}}{2\pi^{2}}\sin^{2}\left(  \frac{\pi}{F_{S}}\right)  , \label{EqA16}%
\end{equation}%
\begin{equation}
a_{3}=\frac{d_{p}}{3\pi}\sin\left(  \frac{3\pi}{F_{S}}\right)  -\frac
{d_{p}^{2}}{2\pi^{2}}\sin\left(  \frac{\pi}{F_{S}}\right)  \sin\left(
\frac{2\pi}{F_{S}}\right)  +\frac{d_{p}^{3}}{6\pi^{3}}\sin^{3}\left(
\frac{\pi}{F_{S}}\right)  , \label{EqA17}%
\end{equation}
etc.

The inverse Fourier transformation of $E_{0}(l,\nu)$,%
\begin{equation}
E_{0}(l,t)=\frac{1}{2\pi}\int_{-\infty}^{+\infty}E_{0}(l,\nu)e^{-i\nu t}dt,
\label{EqA18}%
\end{equation}
gives the solution%
\begin{equation}
E_{\text{out}}(t)=e^{-d_{p}/2F_{S}}\sum_{k=0}^{+\infty}a_{k}E_{\text{in}%
}\left(  t-kT\right)  . \label{EqA19}%
\end{equation}
where $E_{\text{out}}(t)=E_{0}(l,t)$ is the field at the exit of of the medium
and $T=\pi/\nu_{0}$ is a delay time. Maximum amplitude of the prompt pulse
with $k=0$ in Eq. (\ref{EqA19}) decreases according to the equation
$E_{\mathrm{pr}}=e^{-d_{eff}/2}E_{0}$ where effective thickness
$d_{\mathrm{eff}}=d_{p}/F_{S}$ is reduced with respect to $d_{p}$ by a factor
of finesse of the square comb, $F_{S}=\nu_{0}/\delta$. Maximum amplitude of
the first delayed pulse with $k=1$ is described by the equation $E_{1}%
=a_{1}e^{-d_{\mathrm{eff}}/2}E_{0}$ where $a_{1}=\frac{d_{\mathrm{eff}}F_{S}%
}{\pi}\sin\left(  \pi/F_{S}\right)  $. Maximum intensity of this pulse
$I_{1}=\left\vert E_{1}\right\vert ^{2}$ has a global maximum $I_{\mathrm{gl}%
}=\left(  \frac{2F_{S}}{\pi}\right)  ^{2}\sin^{2}\left(  \pi/F_{S}\right)
e^{-2}I_{0}$ for $d_{\mathrm{eff}}=2$ or $d_{p}=2F_{S}$ where $I_{0}%
=\left\vert E_{0}\right\vert ^{2}$. Dependence of $I_{\mathrm{gl}}$ on the
inverse value of finesse, $1/F_{S}$, is shown in Fig. 2(b). From this figure,
it follows that efficiency of the AFC memory, $I_{\mathrm{gl}}/I_{0}$,
increases with increasing finesse. The parameter $d_{p}$ corresponding to this
efficiency also increases according to the relation $d_{p}=2F_{S}$.
\begin{figure}[ptb]
\resizebox{0.6\textwidth}{!}{\includegraphics{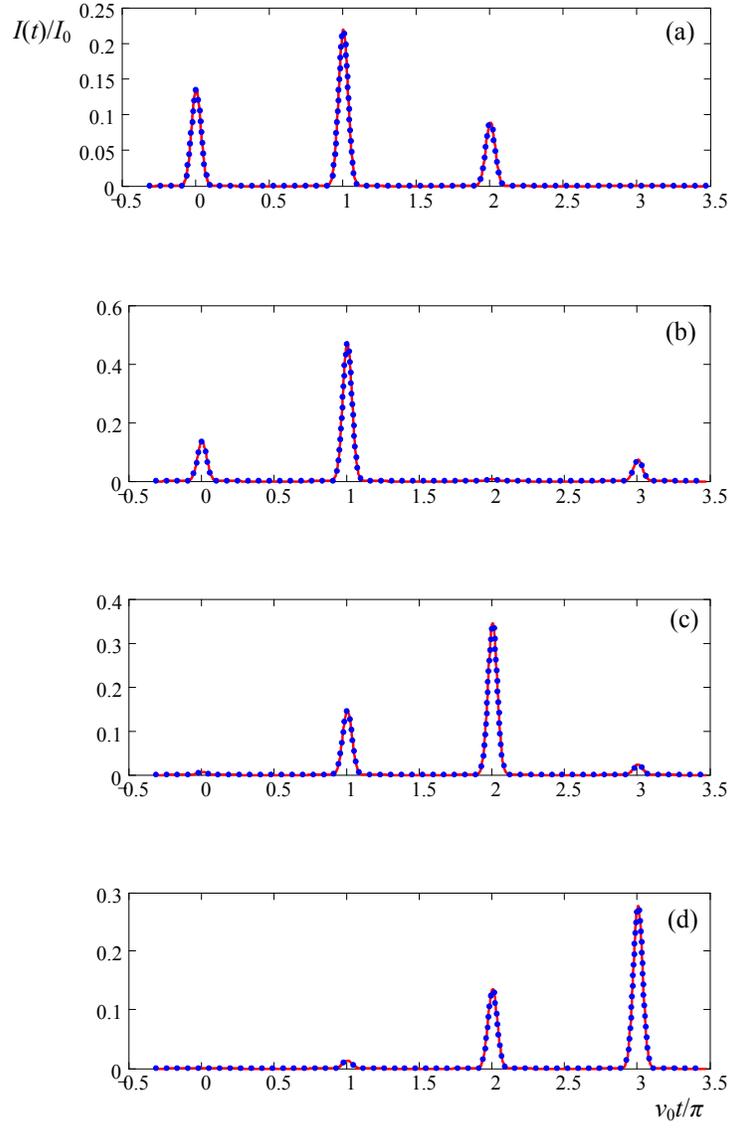}}\caption{Time
dependencies of the intensities of the pulses produced at the exit of the AFC
medium whose parameters are $F_{S}=2$, $d_{p}=4$ (a), $F_{S}=5$, $d_{p}=10$
(b), $F_{S}=5$, $d_{p}=25$ (c), and $F_{S}=5$, $d_{p}=42$ (c). The values of
the peak absorption parameters $d_{p}$ are taken equal to those corresponding
to the global maximum of the first delayed pulse in (a) and (b), the second
delayed pulse (c), the third delayed pulse (d), see Fig. 7 as the reference
for the relevant values of $d_{p}$. Solid red lines are the plots of the
analytical solution Eq. (\ref{EqA19}). Dotted blue lines are the numerical
calculations of the inverse Fourier transformation of the solution
(\ref{EqA9}) for the Gaussian pulse $E_{\text{in}}(t)=E_{0}e^{-\sigma^{2}
t^{2}}$ with $\sigma=5\nu_{0}$. The number of the absorption peaks in the
square AFC is 20 and the frequency integration interval is $(-4\sigma
,+4\sigma)$.}%
\label{fig:6}%
\end{figure}

It is interesting to notice that for a large finesse and the optimal value of
optical thickness, $d_{p}=2F_{S}$, of moderate value the incident field after
passing through AFC medium is mainly distributed between the prompt and first
delayed pulses, see Fig. 6(b). While, for the lowest finesse value $F=2$, the
amplitude of the pulse delayed by time $2T$ (the second delayed pulse) is
comparable with the amplitudes of the prompt and first delayed pulses, see
Fig. 6(a). For essentially larger values of the optical thickness ($d_{p}\gg
1$) the intensities of the delayed pulses are distributed such that delay time
of the pulse with maximum amplitude increases, see Fig. 6(c,d). This simply
follows from the dependence of the intensities of the delayed pulses,
$I_{k}=a_{k}^{2}\exp(-d_{p}/F_{S})I_{0}$, on the optical thickness $d_{p}$,
shown in Fig. 7 for $F=5$ and $k=1$, $2$, and $3$. \begin{figure}[ptb]
\resizebox{0.6\textwidth}{!}{\includegraphics{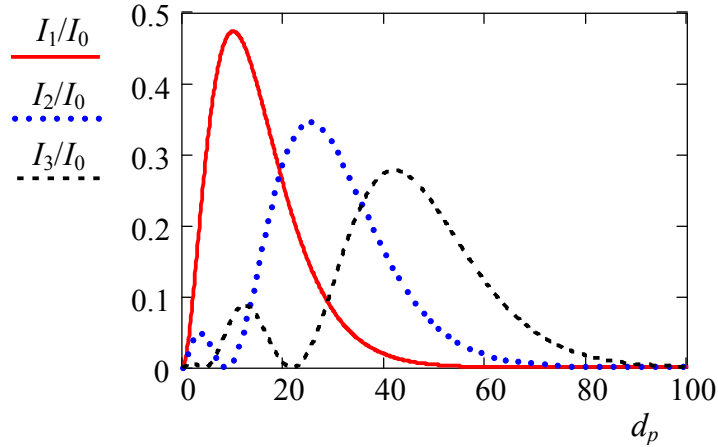}}\caption{Dependencies
of the intensities of the first $I_{1}$ (solid red line), second $I_{2}$
(dotted blue line), and third $I_{3}$ (dashed black line) delayed pulses with
time delays $T$, $2T$, and $3T$, respectively, on the optical thickness
$d_{p}$. Finesse of the square AFC is $F_{S}=5$.}%
\label{fig:7}%
\end{figure}

Exact solution for the harmonic AFC (see Ref. \cite{Shakhmuratov2018}),%
\begin{equation}
E_{\text{out}}(t)=e^{-d_{p}/4}\sum_{k=0}^{+\infty}\frac{(d_{p}/4)^{k}}%
{k!}E_{\text{in}}\left(  t-kT\right)  , \label{EqA20}%
\end{equation}
whose finesse is $F_{H}=2$, shows quite different results for the large
optical thickness. The line, which links maximum amplitudes of the pulses,
forms a bell-shaped envelope, i.e., the energy of the field is smoothly
distributed among the delayed pulses. Numerical analysis shows that similar
results are obtained also for the square AFC for $F_{S}=2$ and large $d_{p}$.
However, after a series of pulses with noticeable amplitudes forming a set
with a bell-shaped envelope, a series of pulse groups with much smaller
amplitudes is formed. This analysis is performed by the numerical calculation
of the coefficients $a_{k}\exp(-d_{p}/2F_{S})$ with the help of equation%
\begin{equation}
a_{k}e^{-d_{p}/2F_{S}}=\frac{1}{2\nu_{0}}\int_{-\nu_{0}}^{\nu_{0}}%
e^{-\alpha_{c}(\nu)l-ik\pi\nu/\nu_{0}}dt. \label{EqA21}%
\end{equation}
\begin{figure}[ptb]
\resizebox{0.6\textwidth}{!}{\includegraphics{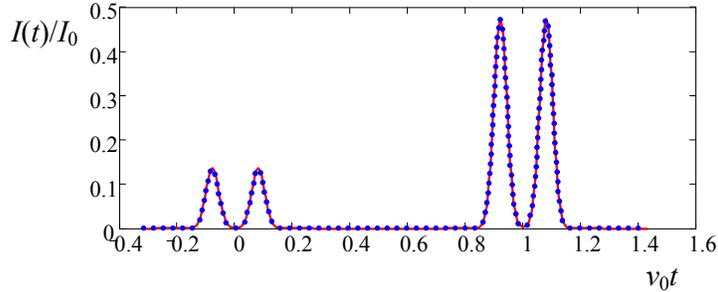}}\caption{Time
dependencies of the intensities of the couple of pulses filtered through the
square AFC comb. The parameters are $F_{S}=5$, $d_{p}=10$, and $\sigma
=7\nu_{0}$. Analytical solution, Eqs. (\ref{EqA23}) and (\ref{EqA24}), is
shown by solid red line. Blue dotted line is the numerical calculation of the
inverse Fourier transformation of the solution (\ref{EqA9}). Frequency
integration interval is $(-4\sigma,+4\sigma)$.}%
\label{fig:8}%
\end{figure}

\section{Appendix B}

In this Appendix the transformation of two closely spaced pulses through the
square AFC comb is considered. The incident radiation consists of two Gaussian
pulses%
\begin{equation}
E_{in}(t)=E_{01}e^{-\sigma^{2}(t+\tau/2)^{2}}+E_{02}e^{-\sigma^{2}%
(t+\tau/2)^{2}+i\varphi}, \label{EqA22}%
\end{equation}
where $E_{01}$ and $E_{02}$ are the amplitudes, $\varphi$ is the relative
phase, and $\tau$ is the time interval between pulses. Below, these pulses
will be denoted as $E_{01}(t+\tau/2)$ and $E_{02}(t-\tau/2)$, respectively.
Since the pulses are weak and do not overlap, in the linear response
approximation one can consider them separately. All second order effects such
as cross-talk of the pulses are neglected.

We consider the square AFC with optimal value of optical thickness, which is
$d_{p}=2F_{S}$. Then, substantial part of the radiation field, filtered
through the AFC, is concentrated in the prompt, $E_{p}(t)$, and delayed,
$E_{d}(t)$, pulses. They are described as follows%
\begin{equation}
E_{p}(t)=e^{-d_{p}/2F_{S}}\left[  E_{01}(t+\tau/2)+E_{01}(t+\tau/2)\right]  ,
\label{EqA23}%
\end{equation}%
\begin{equation}
E_{d}(t)=a_{1}e^{-d_{p}/2F_{S}}\left[  E_{01}(t-T+\tau/2)+E_{01}%
(t-T+\tau/2)\right]  . \label{EqA24}%
\end{equation}
Thus, the delayed pulses have the same relation of the amplitudes and phases
as the couple of the incident pulses. This case is demonstrated in Fig. 8.

\end{document}